\documentstyle[12pt,aasms4,psfig]{article}

\received{: April 99}
\accepted{: June 1999}
\slugcomment{to appear in \it{The Astrophysical Journal}}

\singlespace

\begin{document}

\title{THREE-DIMENSIONAL SIMULATIONS OF JET/CLOUD INTERACTIONS: 
STRUCTURE AND KINEMATICS OF THE DEFLECTED JETS} 

\author{ Elisabete M. de Gouveia Dal Pino$^{1,2}$ }
\altaffiltext{1}{Instituto Astron\^omico e Geof\'{\i}sico, Universidade
de S\~ao Paulo, Av. Miguel St\'efano, 4200, S\~ao Paulo
04301-904, SP, Brasil; 
E-mail: dalpino@iagusp.usp.br}
\altaffiltext{2}{
Astronomy Department,
Theoretical Astrophysics Center,
University of California at Berkeley,
601 Campbell Hall,
Berkeley, CA 94720-3411}

\vskip 2.0 cm


\begin{abstract}

We report the results of three-dimensional smoothed particle
hydrodynamics  simulations of interactions of
overdense, radiatively cooling and adiabatic jets with dense, compact
clouds in frontal and off-axis collisions. 
Calculated for a set of parameters which are particularly appropriate
to protostellar jets, 
our results indicate that the interaction produces important 
transient and permanent effects 
in the jet morphology.

In off-axis interactions, the deflected beam 
initially describes a C-shaped trajectory around the
curved jet/cloud contact discontinuity but the 
deflection angle 
tends to decrease with time as the beam slowly 
penetrates the cloud.  
Later, when the jet has 
penetrated most of the cloud extension, the deflected beam 
fades and the jet resumes 
its original direction of propagation. 
During the interaction, a
weak chain of internal knots develops along the 
deflected beam and the
velocity field initially  has a
complex structure 
that later evolves to a more uniform
distribution. The average 
velocity of the deflected beam is consistent with the 
predicted value given by  
$v_{j}^{\prime} \simeq v_j cos {\theta}$ (where  
$\theta$ is the deflection angle, and $v_j$ is the velocity
of the incident beam). The impact also decreases the beam collimation.
Applied to the context of the protostellar jets, this
morphology and kinematics found for the deflected beam 
is very similar to that observed in some 
 candidate systems
like the 
HH 110 jet which has been previously proposed to be the
deflected part of the HH270 jet.
Our simulations also reveal
 the formation of a $head-neck$ 
bright structure
at the region of impact 
which resembles
the morphology of the HH 110 knot A 
located in the apex of the HH 110 jet where the deflection 
is believed to occur.
All these similarities 
strongly support the proposed jet/cloud interaction interpretation
for this system.
The fact that the deflection angles derived from the simulations
are smaller than that observed 
and the fact that the jet/cloud interaction is 
still taking place  
indicate that the interacting 
cloud in that system must have a radius $R_c \gg R_j$,
where $R_j$ is the jet radius,
as previously suggested, and a 
density ratio between the jet and the cloud 
 $\beta^{2} = n_j/n_c \lesssim 10^{-2}$.

Due to the 
small  size of the clouds 
[with radius $R_c \simeq (1-2) R_j$], the
interactions examined here are very transient 
(with lifetimes  of  few $\sim 10$  to  $\sim 100$ yr  which are
$\ll$ than the typical dynamical lifetimes of 
the protostellar outflows,
$\tau \gtrsim 10^{4}$ yr). Nonetheless, 
they leave important signatures in the surviving outflow. 
The 
left-overs of the cloud
and the knots that are produced in the deflected beam 
are deposited into the working surface
and contribute  
to  enrich the knotty pattern commonly observed in HH objects
behind the bow shocks of protostellar jets.
Also, the collision may partially destroy the
shell at the head producing remarkable asymmetries in the
head region. A
jet undergoing many transient 
interactions with compact clumps along its propagation and lifetime 
may  inject a considerable amount of 
shocked jet  material sideways
into the surrounding
ambient medium and
this process 
may provide a powerful  tool for momentum transfer 
and
turbulent mixing with the ambient medium.


\end{abstract}

\clearpage

\section{Introduction}

The collimated, highly supersonic Herbig-Haro (HH) jets that emerge from protostars in star forming regions propagate through a complex ambient medium composed 
of many dense cloud cores and may eventually collide with some of these
objects.
Although these outflows 
are usually observed to propagate away from their sources 
in approximately straight lines (see e.g., Reipurth 1997 for a 
recent review on protostellar outflows), there are some 
cases where the outflow is observed to be deflected. 
Among these,
there are some examples, like 
HH 270/110 (Reipurth et al. 1996, Rodriguez et al. 1998), 
HH 30 (L\'opez et al. 1995), and the
molecular outflow in L 1221 (Umemoto et al. 1991) 
in which the deflection seems unlikely to
be caused by peculiar changes in the direction of motion of the central
source. Particularly in the first 
case, there seems to be strong 
evidence that the jet deflection is caused by the encounter with a dense 
cloud.

Previous  analytical and numerical work was performed 
to investigate the hydrodynamics of interactions
of shock-fronts and  supersonic winds with interstellar clouds 
(see, e.g., McKee \& Cowie 1975,
Rozyczka \& Tenorio-Tagle 1987, Klein, McKee, \& Colella 1994,
Xu \& Stone 1995, Raga et al. 1998), but all those studies
have focused on the effects of the interaction on the 
structure of the cloud.
The problem of the interaction of an astrophysical jet with a 
rigid surface  has 
been  investigated analytically with some detail by 
Canto, Tenorio-Tagle \& Rozyczka 
(1988, hereafter CTR).  
In the context of extragalactic jets, the well known correlation and 
spatial alignment between radio and 
optical structures in extended extragalactic radio-sources and Seyfert 
galaxies (e.g., McCarthy et al. 1987, Viegas \& de Gouveia Dal Pino 1992) 
have led to a series of analytical 
and numerical studies involving the interaction of 
light jets with ambient clouds (e.g., Higgins at al. 1995,  
Fedorenko \& Courvoisier 1996, Steffen et al. 1997). 

In the context of 
radiatively cooling, heavy jets (i.e., denser than their surroundings)
$~-~$ a picture believed to be consistent with protostellar jets,
the problem of jet/cloud encounters 
and their effects on the jet structure has been discussed by
Raga \& Canto (1995) who focused in the early stages of
the interaction using a simple analytical model and 
two-dimensional simulations involving slab jets impacting a large, 
flat surface
of high density at an arbitrary angle of incidence. 
Subsequently, Canto \& Raga (1996)  and Raga \& Canto (1996) 
have discussed a
 steady-state solution 
based on Bernoulli's theorem for adiabatic 
and radiatively cooling jets  penetrating  into a cloud with
 a plane-parallel  
pressure stratification with exponential profile 
(Canto \& Raga 1996) and  a spherically symmetric 
pressure stratification 
with  power-law  profile (Raga \& Canto 1996). Also,
de Gouveia Dal Pino,
Birkinshaw \& Benz (1996; hereafter GBB96) 
and de Gouveia Dal Pino \& Birkinshaw (1996;
hereafter GB96)
have examined numerically
 the structure and evolution  of jets normally
 propagating  through
stratified environments with different power-law 
pressure distributions.

In the present work, with the help of fully three-dimensional simulations 
(which naturally retain sensitivity to asymmetric effects) 
 we attempt to extend these prior studies by examining
the structure and  evolution of radiatively cooling and adiabatic
jets undergoing frontal and off-axis collisions with compact clouds of 
finite density. 
Although the typical sizes of the cloud cores in 
star forming regions are usually much higher than the jet radius ($R_j$),
with the assumption here of  more compact clumps (with radius
$R_c \gtrsim R_j$) we are able to follow the whole evolution of the 
interacting
system and the deflected beam (whose lifetime time must not much
exceed  
the time it takes for the shock that develops
with the impact to travel over the cloud diameter).   
Besides, with this assumption we can more realistically 
examine, for example,
 the effects of the 
impact of a beam
with the $edges$ of an ambient cloud in an off-axis collision, and also the
effects of the curvature of the cloud in the beam deflection 
$-$ effects that are clearly absent in plane-parallel  analyses.


In \S
2 of this paper, we outline the numerical method and the initial conditions.
 In \S 3, we present the
results of the simulations, and in \S
4 we address the conclusions and the possible
implications of our results for protostellar jets.

\section{The Numerical Model and Setup}

To  simulate the jet/cloud interactions, we have
employed a modified version of our three-dimensional hydrodynamical 
code based on the smoothed particle 
hydrodynamics (SPH) technique (de Gouveia Dal Pino \& Benz 1993;
see also Benz, 1990, 1991, Monaghan 1992, and Steinmetz \& Mueller 1993 for
an overview of the method and a 
discussion of the capabilities and limitations of the SPH).
SPH is a Lagrangean, gridless approach  to fluid dynamics in
which particles 
track the flow and move with it. The code solves
the hydrodynamics equations of continuity, momentum, and energy
explicitly in time, in Cartesian coordinates.
Originally developed to investigate interactions
  between planets and 
 planatesimals (e.g., Benz, Cameron, \& Melosh 1989),
and stellar encounters (e.g., Benz, Bowers, Cameron, \& Press 1990, 
Davies, Benz, \& Hills 1992), 
the code was successively implemented to study
supernova explosions (e.g., Herant, Benz,
\& Colgate 1992), and the structure and evolution of overdense jets
propagating into initially homogeneous ambient media 
(de Gouveia Dal Pino \& Benz 1993, hereafter GB93).
The later version was subsequently modified to investigate
pulsed jets (de Gouveia Dal Pino \& Benz 1994, hereafter GB94);
molecular outflows and related processes of momentum 
transfer between the jet and the molecular environment
(Chernin, Masson, de Gouveia Dal Pino, \& Benz 1994);
jet propagation into stratified environments in star 
formation regions (GBB96 and GB96); and more recently,
the 
effects of magnetic
fields on the structure of overdense radiatively cooling jets
(e.g., Cerqueira,  de Gouveia Dal Pino, \& Herant 1997, Cerqueira 
\& de Gouveia Dal Pino 1999, hereafter CG99). 

 A series of validation tests of the code 
and direct comparisons with results of 
standard Eulerian, grid-based calculations in two and three-dimensions
were successfully performed $-$ our SPH calculations
produce similar results to those of grid-based schemes,
 even considering
a small number of particles, although the  spatial resolution 
may be inferior (with the production of  $smoother $ interfaces) 
due to intrinsic numerical diffusion
(GB93, GB94, Chernin et al. 1994, GB96, CG99, Benz 1990; see also  
Davies et al. 1993 for a detailed comparison of SPH calculations with
 those produced by different finite-difference methods).

In the
present work, we have modified the pure hydrodynamical  version of the 
code  above (see GB93 and GB96) by introducing
a cloud in the homogeneous ambient domain.
Following is a short summary of the key features of the code that
are relevant to this work (for more detailed discussion
of the basic assumptions see the references above).

The computational domain is a 3-D rectangular box 
which represents the ambient medium and has    
dimensions 
$-16R_j \le$ x $\le 16R_j$, $-6R_j \le$ y,z $\le 6R_j$, 
where $R_j$ is
the
initial jet radius (and $R_j$ is the code distance unit).
The Cartesian
coordinate system has its origin at the  center of the box
and the jet  flows through the x-axis, and is continuously 
injected into the 
bottom of the box [at
$\vec{r}=(-16R_j,0,0)$]. Inside the box, the particles are
initially distributed on a Cartesian grid. 
Outflow boundary conditions are
assumed for the boundaries of the box (GB96). 


The particles are
smoothed out by a spherically symmetric kernel function of width $h$.
As in previous work (GB93, GB94, GB96, CG99), the initial resolution, 
as characterized by the initial value of $h$,
was chosen to be $0.4 R_j$ and $0.2
R_j$ for the ambient, and the jet and cloud particles (see below), respectively. 
With this initial particle spacing, the calculations are started with about 
74,000 particles.
Former validation tests (e.g., GB93) have shown that the above choice of 
initial values of $h$ is
more than appropriate to reveal (with accuracy and reasonably good definition) 
the physical properties of the structure of overdense jets. 
(The decrease of these initial values by a 
factor two, for example, enormously increase the number of particles in the
system and thus the required amount of computer time and space, without  significant
improvement to the resolution.) During the evolution of the flow, $h$
is consistently allowed to vary (Benz et al. 1990)
and the resolution is naturally established by the particle
distribution $-$ high-density regions (like shock zones) 
have higher resolution because of the larger concentration of particles,
while low-density regions (like the cocoon that envelopes the 
beam) retain a lower resolution (see below). 
(For a discussion on the basic criteria to monitor accuracy in our SPH code and also in SPH schemes in general,
see also 
Benz 1990, 1991, Monaghan 1992, and references therein.)

As in previous work,  the jet, the cloud, and the ambient 
gas are treated as a fully ionized fluid with an adiabatic index 
$\gamma=5/3$ and an ideal equation of state.
The
radiative cooling, which is due to collisional excitation and 
recombination, is
implicitly calculated using a time-independent cooling function for a
gas of cosmic abundances cooling from $T \simeq 10^6$ to $10^4$ K (the
cooling is set to zero for $T ~{\rm <} ~ 10^4$ K; see GB93, GB96).
The time integration is done using a second order Runge-Kutta-Fehlberg
integrator. The shock waves which arise in the flow are handled by the
usual
Newmann-Ritchmyer artificial viscosity and a link-list
method is used to find the particle's neighbors (e.g., Monaghan 1992).  

An initially isothermal  cloud with a  Gaussian density (and pressure) profile
$$ n_{cl} =  n_a \,  + \,  n_c \, \exp ^{-(\vec{r} - \vec {r_c})^2/\sigma^2}, 
\eqno(1)$$
is 
placed  near the center  of the computational domain, 
where $n_a$ is the ambient number density, 
$n_c$ is the number density in the center of the cloud, 
$ r_c^2 = x_c^2 + y_c^2 + z_c^2 $  is its central coordinate,  and
$\sigma$  is the width of the Gaussian profile which 
we assume to be 
$\sigma$ = 0.75 
$R_c$ in all the simulations, where 
$R_c$ is the initial radius of the cloud. 
For simplicity, the initial temperature of the cloud is
assumed to be the same as that of the surrounding ambient medium
($T_{cl} = T_a$). 

During the  simulations, the cloud 
is held steady by the application of an
appropriate gravitational potential upon its particles. 
Similarly to previous studies of jet propagation
in stratified environments (Hardee et al. 1992, GBB96, GB96),
this external  potential is simply evaluated by assuming that the
cloud is initially in hydrostatic equilibrium so that
$\vec {\nabla } p_{cl} = \rho_{cl} \, \vec {g}$,
where $p_{cl} = n_{cl} \, K \, T_a$  is the thermal pressure in the cloud,
$\rho_{cl} = \bar m \,  n_{cl}$ is the mass density,
and
$\vec {g}$ is the gravitational acceleration. Substituting eq. (1) into 
the equation above and performing the derivation, it yields 

$$ \vec {g} \, =  \, - {2 K T_a \over \bar m \sigma^2} \, 
\,  (\vec {r} - \vec {r_c} ) \, { n_c \exp ^{-(\vec{r}-\vec{r_c})^2/\sigma^2} \over n_a \,
+ \, n_c \exp ^{-(\vec{r}-\vec{r_c})^2/\sigma^2} } 
\eqno(2)$$
where $\bar m \simeq 0.5 m_H$ is the mean mass per particle, 
$m_H$ is the hydrogen mass, and $K$ is the 
Boltzmann constant. 

Since in  the SPH scheme it is trivial to distinguish between particles 
in the jet and those in the ambient medium or  in the cloud, 
it is easy to have the 
external force above computed only over the cloud particles, and
once a particle is swept from the cloud by the impinging jet this force is
no longer computed on it.  During the jet/cloud interaction, this 
force becomes negligible with respect to
the impact forces. Its effects 
on the jet dynamics 
are  also negligible since the jet is assumed to be highly supersonic 
(see below).

The models are parameterized by the dimensionless numbers: 
{\it i}) the density ratio between the jet and the ambient medium,
$\eta=n_j/n_a$; {\it ii}) the
ambient Mach number, $M_a=v_j/c_a$ (where $v_j$ 
is the jet velocity and $c_a$ is the ambient sound speed);
{\it iii}) the jet to
the ambient medium pressure ratio at the jet inlet, $\kappa=p_j/p_a$,
that we assume to be equal to unit; 
{\it iv}) the square root of the density ratio between the jet and the center 
of the
cloud $\beta = (n_j/n_c)^{1/2}$ (see, e.g., Canto \& Raga 1995);
{\it v}) the ratio between the cloud and the jet radius 
$R_c/R_j$;
and {\it vi}) the ratio of
the cooling length in the post-shocked gas behind the bow shock to the
jet radius $q_{bs} = d_{cool}/R_j$ (see, e.g., GB93). 

The major advantages and limitations of our SPH calculations have 
been addressed in previous work (GB93, Chernin et al. 1994, GB96); 
in particular, two points 
should be remarked. 
Firstly, as we mentioned 
above, the 
properties of low-density regions are more poorly sampled
because they do not contain as many particles as a denser region.
However, not having to calculate properties of $ empty $
regions is actually one of the advantages of SPH codes over 
fixed grid based codes. Secondly,
turbulent effects which should exist in these flows
(since the expected Reynolds numbers are very high, $Re > 10^4$; GB93),
are more difficult to examine because the numerical viscosity of the
code may be too dissipative and the initial particle spacing
is large relative to the size of the eddies that may develop in
the flow ($\sim d_{cool}/Re < R_j/10^4$). 
Thus, for this work, we can only consider the 
bulk properties of the jet/cloud  interaction, i.e., over
a size scale larger than that of most of the 
eddies, at which the internal turbulent motions are averaged
out. 

\section{The Simulations}

We have carried out a series of numerical experiments of jet/cloud encounters 
involving clouds with $R_c = (1 - 2) R_j$ and 
 initial jet/cloud density parameter
$\beta \simeq  3.5 \times 10^{-2} - 2 \times 10^{-1}$.
The parameters of the simulations were chosen to resemble typical
conditions found in protostellar jets and their environment. 
We have adopted  an initial 
number density ratio between the jet and the ambient medium
$\eta=n_j/n_a=3$, $n_j=600$ cm$^{-3}$, ambient Mach numbers $M_a =
v_j/c_a = 12 - 24 $ (with $v_j \simeq 200 - 400$ km s$^{- 1}$ 
and 
$c_a = (\gamma K T_a /\bar m)^{1/2} \simeq 16.6 $ km s$^{-1}$
is the ambient
sound speed), and $R_j = 2 \times
10^{15}$ cm.  The corresponding initial jet Mach numbers are
numbers $M_j = v_j/c_j \simeq 20.8 - 41.7 $ 
and 
$c_j = (\gamma K T_j /\bar m)^{1/2}= c_a/(\eta)^{1/2} \simeq 9.6 $ km s$^{-1}$
(this last condition on $c_j$ is due to the assumed pressure equilibrium
at the jet inlet; see \S 2).

The paragraphs below present the results of the simulations performed 
for both off-axis and frontal collisions involving  radiatively
cooling and adiabatic jets, and Table 1 summarizes the values of the
input parameters.

\subsection {Radiatively Cooling Jet/Cloud Off-Axis Interactions} 

Figure 1 shows the results of a jet/cloud off-axis 
interaction in the x-y plane. It depicts the density 
contour in the mid-plane section
and the velocity field distribution 
evolution of a
 radiatively cooling jet which impacts  a dense cloud 
with an initial radius
$R_c = R_j$, 
central coordinates (0, 1.2 $R_j$, 0), and a square root of the
jet to central cloud density ratio 
$\beta  \approx 4 \times 10^{-2}$.
The incident jet has initial 
$\eta =3$  and 
ambient Mach number $M_a=12$ (see Table 1), and 
the cloud has 
central coordinates (0, 1.2 $R_j$, 0).

We find that a double shock pattern develops in the region of impact
 at $t/t_d=2.0$ (where  $t_d=R_j/c_a \simeq $ 38 yr corresponds to the 
transverse jet dynamical time). The incident beam is deflected by 
a  shock nearly parallel to the  surface of the cloud and the 
pressure behind this  induces  a second shock 
which slowly propagates into the dense cloud
 causing an increase in 
its central density by a factor $\sim$ 1.5, from 
1850 $n_a$ before the impact to $\sim$ 2800 $n_a$ at the quasi-steady regime
(see below)
which is attained after the impact (at $t/t_d >$ 4.0).
As expected from previous analytical study (CTR),
due to the highly radiatively cooling regime (which is expected in protostellar 
outflows), 
the angle between the two shocks is very small and they are 
both effectively parallel to the cloud surface at the impact zone. 

Due to the impact, the beam is initially deflected by an angle 
$\theta \simeq 40^o$ but 
this deflection angle tends to decrease with time as the beam partially 
penetrates the cloud and describes a C-shaped trajectory around the
curved jet/cloud contact discontinuity. After $t/t_d = 4.5$, the
bow shock leaves the computational domain and the deflected 
beam propagates at an apparently  $quasi-steady$ $state$ 
regime with nearly constant
deflection angle $\theta \simeq 30^o$ and   velocity field 
distribution. Later on, however, as the jet slowly  excavates a 
way through the dense cloud this regime must have an end
 and the jet will continue
its propagation without significant deflection (see below). 

Weak internal knots develop along the deflected beam as indicated 
in Fig. 1 and also by the
density
and pressure profiles across the flow depicted in Fig. 2, for 
$t/t_d =$ 4 and 6.5. In the first stages of the interaction, 
the velocity field along the deflected beam is somewhat 
complex with velocity fluctuations along the 
ejected parts of the beam which
are correlated with the positions of the knots.
At $t/t_d =$ 4, it varies from $\sim$ 11.8 $c_a$ near the region of 
impact, to $\sim$ 12.3 $c_a$ at x = 4 $R_j$,
$\sim$ 10.0 $c_a$ at x = 5 $R_j$, and
$\sim$ 10.5 $c_a$ at x = 10 $R_j$.
Later on, the velocity 
field tends to become more uniform and slightly
decreases with distance from the impact region. At $t/t_d=$ 6.5, 
it  decreases from 
$\sim$ 11.8 $c_a$ near the region of impact to $\sim$ 10.5 $c_a$
in the 
middle and $\sim$ 10 $c_a$ at the end of the outflow. 
This corresponds
to an average velocity for the deflected beam 
$v_{j}^{\prime} \simeq 10.8$ $c_a$, which is
 compatible with 
the expected value (see e.g., CTR)
$$v_{j}^{\prime} \simeq v_j cos{\theta},
\eqno(3)$$
which gives $v_{j}^{\prime} \simeq$ 10.4 $c_a$ 
(for  $v_j =$ 12 $c_a$ and $\theta \simeq 30^o$).

Before the impact, the bow shock propagates downstream with a
speed (GB93)
$v_{bs} \simeq v_j/[1 + {(\eta \alpha)}^{-1/2}] \simeq 8$,
and after the impact 
$v_{bs}^{\prime} \simeq 5.5 $,
which roughly agrees with the predicted estimate
$v_{bs}^{\prime} \simeq v_j^{\prime}/[1 + {(\eta \alpha)}^{-1/2}]\simeq 5$,
 for a measured 
$\alpha = (R_j/R_h)^2 \simeq 1/4$  from the simulation (where 
$R_h$ is the radius at the
jet head which is initially equal to $R_j$ and increases 
by a factor about two  
after the impact, see Fig. 1).
The impact reduces the jet collimation, as expected
(e.g., CTR, Raga \& Canto 1995), and the reflected
beam has an initial opening angle $\psi \simeq 30^o$ 
which reduces to $\psi\simeq 20^o $, as 
indicated by the density contour maps of Fig. 1,
after $t/t_d$= 4. 

We note that the  deflected jet fades as it propagates downstream at a 
distance 
$\sim 10 R_j$ and this fading can be testified by the density
and pressure profiles across the flow depicted in Fig. 2, for 
$t/t_d =$ 6.5. 
This fading is in part caused by a drastic spreading of the jet material
in the other directions mainly due to the more frontal interaction
 that
the beam experiences with the cloud in the x-z plane (see Fig. 3).
For comparison, Figure 4 shows the result of a jet/cloud 
interaction for a system
 with  the same initial conditions as those in Fig. 1 except that  
the cloud is positioned off-axis also in the x-z plane, i.e.,
it  has central coordinates (0, 1.2 $R_j$, 1.2 $R_j$) and is located 
in one of the quadrants of the y-z plane. In this case, the jet 
effectively impacts a smaller section of the cloud and the collision 
is therefore  much
weaker $-$ since the density in the cloud decreases from the center to
the edges according to Eq. (1), the effective density parameter  $\beta$ 
is larger by almost an order of magnitude than that of Fig. 1
($\beta \simeq 2 \times 10^{-1}$). As a consequence, 
the deflection angle of the 
beam is smaller (at the time depicted, 
$\theta \simeq 16^o$  
against 
$\theta \simeq 30^o$ in Fig. 1). In Fig. 4, we also note that
the interaction between the beam and the cloud is almost completed. 
The jet has
already 
penetrated most of the cloud extension and is resuming
its original direction of propagation. Consistently, the deflected jet 
of Fig. 4 has a propagation velocity 
($v_{j}^{\prime} \, \simeq 11.5 c_a$
$ \simeq \, v_j \, cos {\theta} $)
which is larger than that 
of Fig. 1, and is fading earlier as indicated by the density
contour map (the jet density decreases to less than  $n_a$ above
5 $R_j$).

We can estimate a lower limit for the survival
  time $t_c$ of the jet/cloud interaction and the 
complete depletion of the deflected beam
by evaluating the time that the second shock
takes to travel over the cloud diameter $d_c = 2 R_c = f R_j$
(where $f > 0$ is a multiple factor of the jet radius). One-dimension
momentum flux conservation argument and the resulting shock 
geometry for a radiatively  cooling jet interacting with a
cloud of constant density and a density parameter
$\beta = (n_j/n_c)^{1/2} \ll 1$ yields a cloud shock speed
$v_{cs} \simeq \beta v_j \sin {\theta} $
(see e.g., CTR, Raga \& Canto 1995),
and $t_c \gtrsim d_c/v_{cs}$
or 
$$\frac{t_c}{t_d } \gtrsim \frac{f}{M_a \beta \sin {\theta}}
\eqno(4)$$
For the deflected jet of Fig. 1 this  equation gives
$t_c/t_d \gtrsim $ 7.5 (for an average  $\theta = 34^o$ and $f = $2),
so that after this time,
we would expect the incident beam to have 
penetrated the cloud almost completely thus suffering no more deflection.
 This time is, however, just a lower limit because 
as the jet penetrates the cloud both the 
deflection angle and   the density parameter $\beta$ decrease
(due to the compression of the cloud) thus
increasing $t_c$ in the equation above. 
For  the jet in Fig. 4 we estimate 
$t_c/t_d >$ 3.

Figure 5 depicts the mid-plane density contour and the 
velocity field distribution of five radiative cooling 
jets interacting  with clouds with different initial values of
jet/cloud density parameter $\beta$.
All the inicident jets have the same initial conditions as in
Fig. 1.
The top jet is propagating into a  homogeneous 
medium and thus suffers no deflection. The four other
systems have 
$\beta \simeq  2 \times 10^{-1}$, $7 \times 10^{-2}$, 
$4 \times 10^{-2}$ (as in Fig. 1), and
$3.5 \times 10^{-2}$
from top to bottom, respectively.
As expected, the angle of deflection increases 
with  the increase of the density of the cloud (smaller $\beta$) 
and the 
larger the angle the slower the propagation speed of the
deflected beam. 
Likewise, the interaction time increases with
decreasing $\beta$ (Eq. 4) and  the jet with 
$\beta \simeq 2 \times 10^{-1}$ (Fig. 5, second panel), for example,
  has already practically 
resumed its initial direction of propagation after partially 
destroying the cloud, while the jet with  
$3.5 \times 10^{-2}$ (bottom) is still interacting 
with   the cloud.
Equation (3) gives the following lower limit interaction times:
$t_c/t_d >$ 0, 2.5,  5, 7.5, and 8.5, respectively. (We note, 
however, that Eq. (3) is valid essentially for 
$\beta \ll 1$ so that the estimated values above for
$t_c$ for the
systems with larger $\beta$ are less reliable).

Figure 6 shows the evolution of an interacting system 
with initial conditions similar to those of  Figs. 1 and 5
but the cloud  has now a radius twice as bigger
($R_c = 2 R_j$) and an  effective density parameter 
$\beta \approx 6 \times 10^{-2}$ (which is comparable to that of the
 third jet of 
Figure 5).
The larger cross section of the obstacle causes the deflection angle of 
the beam  to be larger and the interaction time much longer
(Eq. 4 gives 
$t_c/t_d >$ 9.3, for a 
cloud diameter $d_c = 4 R_j$, or f = 4). 

The compressing shock increases the central density of the 
cloud to a maximum factor $\simeq$ 1.4 and
the deflection angle of the beam 
varies  from 
$\theta \simeq 45^o$ (at the  initial impact at $t/t_d =$ 2)
 to $\theta \simeq 40^o$ after $t/t_d = $ 3.5
(against $\theta \simeq 25^o$ in the third jet of Fig. 5, and
$\theta \simeq 35^o$ in the jet of Fig. 1 at the same 
evolution time). Consistent with these deflection angles (Eq. 3), the
average velocity of the deflected beam varies from 
$v_{j}^{\prime} \simeq$ 9 to 7 $c_a$,
and the bow shock velocity decreases from $\simeq$ 8 $c_a$ before the
impact to  $\simeq$ 5 $c_a$ after it.
The opening angle of the deflected
beam varies from $\psi \simeq 35^o$ 
to $\psi\simeq 25^o $, as 
indicated by the density contour maps of Fig. 6.
The density contour maps also indicate the formation of 
some bright knots along the deflected beam and bow shock.
Those knots have densities $n_k/n_a \simeq$ 5 to 15 and
were originally  produced in the region of impact at the
contact discontinuity and carried downstream by
the deflected jet.
At $t/t_d =$ 7, the density in the deflected beam decreases
from 
$ n_j^{\prime}/n_a \simeq 5$ near the impact to less than  1 above
$\sim 5 R_j$. 

\subsection {Jet/Cloud Off-Axis Interactions for Jets with Different 
$M_a$}

Figure 7 depicts two interacting systems with
different initial ambient Mach numbers, 
$M_a =$ 12 (top) and 18 (bottom). Both jets have
an initial  density parameter
$\beta \simeq 3.5 \times 10^{-2}$ and
the other initial conditions are the same as  in Figs. 1 and 5.
 (Note that the  jet with $M_a$ = 12 is the same jet 
in the bottom panel 
of Fig. 5.)
The jet with larger $M_a$ interacts faster with the cloud (Eq. 4) 
and thus fades
earlier too
($t_c/t_d >$ 6 and 8.5, for the $M_a =$ 18 and 12, respectively). 
At
the time depicted in Fig. 7, the $M_a = 18$ jet is already 
starting resuming 
its original direction.
Initially, both systems have similar deflection angles
($\theta \simeq 40^o$) but because of the larger interaction rate of
the $M_a=$ 18 jet, at the time depicted its angle has become
smaller ($\theta \simeq 35^o$ in the 
$M_a =$ 18, against 
$\theta \simeq 40^o$ in the $M_a= $12 
jet).
The density in the deflected beam decreases from 
$n_j^{\prime}/n_a \simeq $ 8 near the impact region, to
less than unity above a distance $\sim 5 R_j$ in the
$M_a =$ 18 jet, while in the  $M_a =$12 it decreases 
from $n_j^{\prime}/n_a \simeq $ 11 near the impact region, to
$\simeq $ 2.4 at $\sim 5 R_j$, and   
less than unity above $\sim 8 R_j$.
 
Figure 8 shows an example of an even higher Mach
 number jet ($M_a = $ 24) after it had 
impacted a cloud (in an off-axis collision)
with a relatively large density parameter
($\beta \simeq 2 \times 10^{-1}$ 
which is comparable to
that of the second jet in Fig. 5).
(The counterpart of this jet  propagating
into an initial homogeneous environment can 
be found, e.g., in GBB96, and Cerqueira, de Gouveia 
Dal Pino \& Herant 1997.)
In the time depicted in the figure, the interaction has already
finished and the jet has completely resumed its 
original propagation direction but the interaction left some 
interesting signatures 
in the system. We see that the remains of the cloud are still 
present at x = 0 and have been involved by the bow  
shock structure. Also,
part
 of the dense shell that developed at the
head of the beam from the cooling of the shocked jet material
 has been
detached from the head by the collision 
and left behind in the cocoon causing a 
remarkable asymmetry in the jet head region. 

\subsection {Frontal Jet/Cloud Interactions}

Figure 9 shows an example of a strong frontal impact of a jet with
a dense cloud with a density 
parameter $\beta \simeq 7 \times 10^{-2}$. The 
initial conditions in this system are the same as in the
third jet of Fig. 5 except for the location of the
cloud which  now has coordinates  (0, 0, 0).
The frontal collision is obviously much stronger and
the compression increases the density of the cloud 
by a factor $\sim$ 2.5
(from $n_c/n_a \simeq$  700 before the impact, to 
 $n_c/n_a \simeq$  1800 after it).
The jet splits into two  beams on either side  of the cloud 
with equal deflection angle
$\theta \simeq 45^o$ and a  double bow shock structure (or double lobes) develops.
By the time the bow shocks leave  the computational 
domain ($t/t_d \simeq $ 6) most of the cloud has been destroyed by the interaction
and the  double deflected beam has almost faded, although the jet
has not yet completely resumed its original direction. 
The propagation velocity of the deflected beams varies 
from 7 to 10 $c_a$ along the flow, 
and the average density in the post-impact beam 
is $\sim$ 4 $n_a$.
Simulations involving weaker frontal interactions with less dense clouds
have shown that the beam easily  sweeps the cloud material to
the working surface causing it to become more knotty and wider.

\subsection {Adiabatic Jet/Cloud Interactions}

Figure 10 shows the density in the mid-plane section and the 
velocity field distribution of an  adiabatic jet which is 
interacting with a (adiabatic) cloud in an off-axis collision
after it had propagated over a distance $\sim 30 R_j$.
The initial conditions are the same as in Figure 1. 
The interaction 
causes a compression in the interacting cloud of the same 
amount  as that produced by the radiatively cooling jet
(by a factor $\sim $ 1.5) but the  adiabatic jet
penetrates less deeply  into the cloud describing a more
pronounced C-shaped trajectory around the contact discontinuity.
The deflection angle is in turn larger 
($\theta \simeq 40^o$ against 
$\theta \simeq 35^o $ in the radiatively cooling jet at
$t/t_d =$4) and so the opening angle
(as  indicated by the density contour maps $\psi \simeq 35^o$ 
in the adiabatic jet and 
$\psi \simeq 16^o$ in the radiatively cooling counterpart in Fig. 1). 
This result is consistent with previous analytical 
and two-dimensional numerical studies of the interaction of 
adiabatic and  radiatively cooling jets with 
plane-parallel and spherically stratified  clouds (Canto \& Raga 1996,
Raga \& Canto 1996).
At the time depicted, the adiabatic jet is trying  to resume 
its original direction causing the fading of the deflected beam
(the density in the deflected beam decreases from
$n_j^{\prime}/n_a \simeq$ 9.5 near the impact region to less than unity at a 
distance
$\sim 4 R_j$ in the adiabatic jet, while it decreases from 
$n_j^{\prime}/n_a \simeq$ 11.5  to 1 at a  distance
$\sim 10 R_j$ in the cooling jet).
The average velocity in the adiabatic and radiatively cooling 
deflected beams
are of the same order 
$v_j^{\prime} \simeq$ 11 $c_a$.

\section{Discussion and Conclusions}

We have presented the results of fully three-dimensional simulations of
 overdense, radiatively cooling and adiabatic jets colliding  
with dense, compact 
ambient clouds. Frontal and off-axis collisions were examined.
Evaluated for a set of parameters which are particularly appropriate
to protostellar jets [with initial density ratios between the jet and the
ambient medium $\eta \approx 3$, 
ambient Mach numbers $M_a \approx 12-24$, 
and jet/cloud density parameters 
$\beta = (n_j/n_c)^{1/2}  \simeq  3.5 \times 10^{-2} - 2 \times 10^{-1} $], 
our results indicate 
that important transient and also permanent effects 
may occur on the jet as  a
consequence of the interaction.
Our main results can be summarized as follows.

1. As in previous analytical and
two-dimensional numerical study (CTR, Raga \& Canto 1995), 
we find that the primary effect of a radiatively cooling jet/cloud
collision is to deflect the beam by 
a shock to a direction
initially nearly parallel to the surface of the cloud at the 
impact region. 
A secondary shock, which is induced by the increased pressure behind the
first,
slowly propagates into the dense cloud
causing an overall increase in 
its density. 

2. The deflected beam initially describes a C-shaped trajectory around the
curved jet/cloud contact discontinuity but the 
deflected angle 
tends to decrease with time as the beam slowly 
penetrates the cloud. 
Later, when the jet has 
penetrated most of the cloud extension the 
deflected beam fades and the jet tends to resume 
its original direction of propagation. Due to the 
small size of the clouds 
[with radius $R_c \simeq (1-2) R_j$], the lifetimes of
 the
interactions deduced   from the simulations
are only    $\sim$ few 10  to $\sim$ few 100 yr (for jets 
with $M_a= $ 24 - 12) but they are longer than the predicted 
time from analytical 
modeling (Eq.3;
see also Raga \& Canto 95). 

3. During the interaction, weak internal knots develop along the deflected beam. 
The
velocity field initially  has a
complex structure with  variations along the flow 
that later evolves to a more uniform
distribution. At the region of impact the velocity
is the order of the incident jet velocity, but the average 
velocity of the deflected beam is compatible with the predicted value
$v_{j}^{\prime} \simeq v_j cos {\theta}$, where 
$\theta$ is the deflection angle, and $v_j$ is 
the velocity of the incident beam (e.g., CTR).
The impact also increases the jet opening angle, as expected.

4. Jets in off-axis collisions with clouds with different density
parameter $\beta$ result different deflection angles 
and interacting times $-$
the larger the density of the  cloud  (the smaller the $\beta$)
the 
larger the angle and the longer the interaction time (Fig. 5).
The increase in the radius of the cloud also
makes the deflection angle and the interaction time larger,
while the  
increase in the incident jet Mach number, $M_a$, 
naturally decreases the time of the  interaction. 

5. Frontal collisions with very dense clouds, 
although they are expected to be even rarer,
may also produce peculiar transient features.
They are naturally stronger and faster than off-axis
interactions with similar initial conditions.
Such interactions cause the splitting of the jet 
into two  beams 
which produce  a  double bow shock structure on  either side of the cloud.
This is, however, a very transient feature and thus highly unlikely to
be observed. Weaker frontal interactions, on the other hand, do 
not produce a jet splitting and most of the 
cloud material is simply swept to the working surface at the jet head.

6. Adiabatic jets interacting with clouds in off-axis collisions
penetrate less deeply  into the cloud and describe a more
pronounced C-shaped trajectory around the contact discontinuity.
The deflection angle is in turn larger 
than that in its radiatively cooling counterpart.
This result is consistent with previous analytical 
and two-dimensional numerical studies involving jet 
interactions with plane-parallel and spherically 
stratified obstacles
 (Canto \& Raga 1996, Raga \& Canto 1996).

The basic features found above in the deflected beam, 
such as the decrease in the
jet velocity and the increase in the opening angle 
with respect to the incident beam,
have been detected in the HH 110 jet (Reipurth et al. 1996,
Rodriguez et al. 1998) which is
possibly the most convincing example, 
among the  protostellar jets, of beam 
deflection by interaction with an ambient cloud.
As stressed by those authors, no driving source has been 
detected at the apex of the HH 110 jet and it seems to be
the deflected part of the fainter HH 270 jet.
They have proposed that the deflection is caused
by an interaction of the jet with a dense cloud core 
with $\beta \simeq 0.03 - 0.3$. Since this
system seems to lie close to the plane of the sky we may
directly compare its morphology with the results of the simulations.
Although the measured deflected angle ($\sim 58^o$)
 is larger than those obtained in our simulations,
the $head-neck$ bright structure we see in the density contour
maps of the off-axis simulations at the region of impact
in  strong interactions (see Figs. 1, 5, 6, and 7),
is remarkably similar to the morphology of the HH 110 knot A 
located at the apex of the HH 110 in the region
where the deflection of the HH 270 jet is believed to occur
(see Fig. 6 of Reipurth et al. 1996). 
Besides, the HH 110 flow is observed to have a well collimated
chain of knots near the apex and then to widen in a cone of
large opening angle ($\psi \simeq 12^o$), until it fades in a
final curve (Reipurth et al. 1996). 
All those morphological characteristics are 
clearly detected in our off-axis simulations, and
the wide range of proper motions measured for the knots
in HH 110 is compatible with the complex velocity 
structure  found in the deflected beams.
Furthermore, the estimated average velocity of HH 110 is consistent
with the $v_{j}^{\prime} \simeq v_j \, cos {\theta}$ relation (Reipurth et al. 
1996). 
All these similarities 
strongly support the proposed jet/cloud interaction interpretation
for the HH 110/HH 270 system.
The fact that the deflection angles derived from the simulations
are smaller and the opening angles are larger than those observed 
in the HH 270/HH 110 system [even for the simulations 
involving an incident jet with $v_j \simeq$ 300  km s$^{-1}$ 
(or $M_a = $ 18) which is the order of that  inferred from observations] 
and the fact that the jet/cloud interaction is probably
still taking place (thus requiring an interacting time $\simeq$
the dynamical time of the deflected jet),
indicate that the interacting 
cloud in that system must have a radius $R_c \gg R_j$,
as suggested by Reipurth et al. (1996), and 
a density parameter $\beta \lesssim$ 0.1, as
indicated by the simulations.

Applied to the general context of 
the protostellar jets, 
the interactions examined here are essentially transient 
processes and thus the probability of they being observed  
must be very small,
since the typical dynamical lifetimes of the observed outflows
($\tau  \gtrsim 10^{4} $ yr; e.g., Bally \& Devine 1997) 
are much larger than the inferred survival times of the jet/cloud
interactions.
Nonetheless these interactions may leave a variety of interesting
more permanent features  imprinted in the remaining outflow.
For example, we have found that after a weak interaction
of only few decades 
with a compact cloud (with density parameter $\beta \simeq $ 0.2)
the  $M_a =$ 24  jet has retained in
its working surface 
the remains of the cloud, and
some fragments of the
dense 
shell have been detached from the head during the collision
producing remarkable asymmetries in the beam (Fig. 8).
Weak interactions are also able to produce some 
wiggling in the deflected beam (Figs. 5 and 8) but this feature 
will last not much longer than the interaction.
The production of jet wiggling by the interaction 
with a plane-parallel stratified environment was also
reported in  previous numerical studies (GBB96; GB96).

The simulations also indicate that before a deflected 
beam  fades it may have time enough to 
produce and deposit some knots into the 
working surface at the jet head (see, e.g.,  Fig. 6)
which may contribute 
to  enrich  and enlarge the knotty pattern 
behind the bow shock.
Since this clumpy structure resembles the knotty pattern 
commonly observed in HH jets, this result suggests that 
transient jet/cloud interactions may also play an important role 
in the formation of these HH structures. 

Finally, we should also note that a jet undergoing many transient 
interactions with compact clumps along its propagation and lifetime 
may  inject  a considerable amount of 
shocked jet  material sideways
into the surrounding
ambient medium over a transverse extension  which will depend on the deflection
angle of each interaction
 (Figs. 6 and 7). This process 
may be therefore,  a powerful  tool for momentum transfer 
and
turbulent mixing with the ambient medium  $-$ a process that
may help to feed the slower and wider molecular outflows often associated 
with protostellar jets (see e.g., Raga et al. 1993, Chernin et al. 1994,
Cabrit et al. 1997).

\acknowledgements
This work was partially supported by the
Brazilian agencies FAPESP and CNPq.
The author is indebted to the referee Paul Wiita  for
his fruitful comments. 
The author also would like to acknowledge the kind 
hospitality of the Star Formation Group of the 
Astronomy Department of the 
University of California at Berkeley where most of this 
work was done and also relevant and clarifying  discussions 
with Frank Shu and Bo Reipurth. Technical support from A.H. Cerqueira
is also acknowledged.

\newpage

\centerline {\bf TABLE}

\begin{table}[ht]

\caption{The models and their initial physical parameters.}

\vspace{10pt}
\centering
\footnotesize
\tabcolsep 10pt
\begin{tabular}{ c c c c c c c c } \hline \hline 
Figures & $M_a$ & $M_j$ & $\eta$ & $\beta$ & $R_c/R_j$ & Collision & Cooling \\ \hline
1, 2, 3, 4, 5 & 12 & 20.8 & 3 & 4 $\times 10^{-2}$ & 1 & off-axis & yes \\
5 & 12 & 20.8 & 3 & $  \sqrt {\infty} $  & 1 & $ - $ & yes \\
5 & 12 & 20.8 & 3 & 2 $\times 10^{-1}$& 1 & off-axis & yes \\
5 & 12 & 20.8 & 3 & 7 $\times 10^{-2}$ & 1 & off-axis & yes \\
5, 7 & 12 & 20.8 & 3 & 3.5 $\times 10^{-2}$  & 1 & off-axis & yes \\
6 & 12 & 20.8 & 3 & 6 $\times 10^{-2}$ & 2 & off-axis & yes \\
7 & 18 & 31.3 & 3 & 3.5 $\times 10^{-2}$  & 1 & off-axis & yes \\
8 & 24 & 41.7 & 3 & 2 $\times 10^{-1}$ & 1 & off-axis & yes \\
9 & 12 & 20.8 & 3 & 7 $\times 10^{-2}$ & 1 & frontal & yes \\
10 & 12 & 20.8 & 3 & 4 $\times 10^{-2}$ & 1 & off-axis & no \\
\hline \hline
\end{tabular}
\tabcolsep 6pt
\end{table}

\newpage

\centerline {\bf FIGURE CAPTIONS}

\noindent {\bf Figure 1:}  Mid-plane density contour (left)
and  velocity field distribution (right)
evolution of a
 radiatively cooling jet interacting with a cloud 
with radius
$R_c = R_j$, central coordinates (0, 1.2 $R_j$, 0), and a
Gaussian density (and pressure) profile of width 
$\sigma=0.75$ $R_j$.
The initial
conditions for the jet are: $\eta=n_j/n_a=3$, $n_a=200$ cm$^{-3}$, 
$M_a=12$, $v_j
\simeq 200$ km s$^{-1}$, 
radiatively cooling length parameter
behind the bow shock 
$q_{bs} \simeq 0.5$ and behind the
 jet shock 
$q_{js} \simeq \eta^{-3}q_{bs} \simeq 2 \times 10^{-2}$ (GB93). 
The square root of the density ratio between the jet and the center of the cloud is
$\beta = (n_j/n_c )^{1/2} \simeq 4 \times 10^{-2}$.
The times depicted are:
 $t/t_d=$  2.0, 3.0, 4.0, 5.0 and 6.5  ($t_d = R_j/c_a \simeq 38$ years). 
 The distances are in units of $R_j= 2 \times 10^{15}$ cm and the jet
was injected at x $\simeq$ -16 $R_j$. 
The density lines are separated by a factor of 1.3 in the 
second and third 
panels, and 1.2 in the rest.
The density scale covers the range, from top to bottom:  
$\sim$ (4$\times 10^{-1}$   -  $2.8\times 10^3$) $n_a$,
$\sim$ ($ 10^{-1}$   -  $5.0\times 10^3$) $n_a$, 
$\sim$ ($ 10^{-1}$  -  $3.6\times 10^3$) $n_a$, 
$\sim$ (4$\times 10^{-1}$  -  $2.9\times 10^3$) $n_a$, and
$\sim$ (4$\times 10^{-1}$   -  $2.9\times 10^3$) $n_a$.

\noindent {\bf Figure 2:} Density (solid line)  and pressure
(dashed line) profiles across the flow  
of Fig. 1 at different 
positions along the flow at $t/t_d \simeq $ 4.0 (top) and 6.5 (bottom).
The peaks on the left side of the
jet axis trace the propagation direction of the deflected beam. 
The density and  pressure scales can be calibrated
using the marker in the top region of the plot (the 
marker for the density corresponds to $\sim 1.7 n_a$,
and the marker for the pressure corresponds to
$\sim 3.5 p_a$, where $p_a \simeq 9.3 \times 10^{-10}$ dyn cm$^{-2}$). 
The density and pressure are very high at the impact region 
(at x= 0)
and have been clipped to highlight
the low level features in the deflected beam.

\noindent{{\bf Figure 3:}  The jet/cloud system of Fig. 1 
seen in the x-z plane at 
$t/t_d =$ 6.5. 
Due to the position of the cloud, which has central coordinates
(0, 1.2 $R_j$, 0), the jet/cloud collision seen from this plane 
looks $frontal$.

\noindent{{\bf Figure 4:} 
Mid-plane density contour (top) and velocity field distribution
(bottom) of a 
 radiatively cooling jet interacting with a cloud 
with radius
$R_c = R_j$ and  central coordinates (0, 1.2 $R_j$, 1.2 $R_j$)
at a time  $t/t_d =$ 6.5.
 The other initial conditions  are the
same as in Fig. 1. The collision is this case is $off-axis$ 
in both planes x-y and x-z.
In the contour plot, the density lines  are separated 
by a factor 1.2 and
the density scale covers the range 
$\sim$ ($4 \times 10^{-2}$  -  $2.9\times 10^2$) $n_a$.

\noindent{{\bf Figure 5:}  Mid-plane density contour (left) and 
velocity field distribution (right) of five radiatively cooling 
jets interacting  with clouds with different initial 
density parameter $\beta$ after they have propagated 
over a distance $\approx 30 R_j$ (corresponding to  $t/t_d = 3.5$ 
for the first panel and $t/t_d = 4$ for the rest).
From top to bottom: $\beta \simeq \sqrt {\infty} $, $2 \times 10^{-1}$, 
7 $\times 10^{-2}$, 4 $\times 10^{-2}$,
  and
3.5 $\times 10^{-2}$.
The other initial conditions are the same as in Fig. 1.
The deflection angle in each case, from top to 
bottom, is:
$\theta \simeq  0^o$, $15^o$, $25^o$, $35^o$, and $40^o$,
respectively. 
The density scale in the contour maps 
covers the range, from top to bottom:  
$\sim$ (2 $\times 10^{-1}$  -  $8.7\times 10^1$) $n_a$, 
$\sim$ (4 $\times 10^{-1}$   -  $7.4\times 10^1$) $n_a$,
$\sim$ (5 $\times 10^{-1}$    -  $9.5\times 10^2$) $n_a$, 
$\sim$ ( $ 10^{-1}$   -  $3.6\times 10^3$) $n_a$, and 
$\sim$ (7 $\times 10^{-1}$   -  $5.3\times 10^3$) $n_a$. 

\noindent{{\bf Figure 6:}  Mid-plane density contour (left)
and  velocity field distribution (right)
evolution of a
 radiatively cooling jet interacting with a cloud 
with radius
$R_c = 2 R_j$ and  central coordinates (0, 1.2 $R_j$, 0).
The density parameter is
$\beta  \approx 6 \times 10^{-2}$ (which is comparable to that of the
 third jet of 
Figure 5).
The other
 initial conditions are the same as in  Figs. 1 and 5.
The times depicted are:
 $t/t_d=$  2.5,  4.5, 5.5 and 7.0. 
The density scale in the contour maps 
covers the range, from top to bottom:  
$\sim$ (3$\times 10^{-1}$    - $2.2\times 10^3$) $n_a$, 
$\sim$ (2$\times 10^{-1}$   - $1.8\times 10^3$) $n_a$,
$\sim$ (2$\times 10^{-1}$   - $1.6\times 10^3$) $n_a$, and 
$\sim$ (2$\times 10^{-1}$   - $1.4\times 10^3$) $n_a$. 

\noindent{{\bf Figure 7:} 
Mid-plane density contour  and 
velocity field distribution of two radiatively cooling 
jet/cloud systems  with  different initial jet Mach numbers
$M_a$ = 
 12 (top) and 18 (bottom), after they have propagated 
over a distance $\approx 30 R_j$ at $t/t_d = 4$.
 Both jets have
an initial  density parameter 
$\beta \simeq 3.5 \times 10^{-2}$ and
the other initial conditions are the same as  in Fig.1.
 (Note that the  jet with $M_a$ = 12 is the same of the 
bottom panel 
of Fig. 5.) 
The density scale in the contour maps 
covers the range:  
$\sim$ (7$\times 10^{-1}$  - $5.3\times 10^3$) $n_a$ (top), and
$\sim$ (8$\times 10^{-1}$  -  $5.8\times 10^3$) $n_a$ (bottom).
 
\noindent{{\bf Figure 8:} Mid-plane density contour (top) and 
velocity field distribution (bottom) of an $M_a$ = 24  radiatively cooling 
at $t/t_d = 1.7$ after interacting  with a cloud 
with radius
$R_c = R_j$ and central coordinates (0, 1.2 $R_j$, 0).
The square root of the jet to central cloud density ratio is
$\beta  \approx 2 \times 10^{-1}$ and the
initial
conditions for the jet are: $\eta=n_j/n_a=3$,  
 $v_j
\simeq 400$ km s$^{-1}$, $q_{bs}\simeq 8$, and $q_{js} \simeq 0.3$.
The density lines are separated by a factor of 1.2 and
the density scale 
covers the range:  
$\sim $ (2$\times 10^{-1}$   -  $2.4\times 10^2$) $n_a$.

\noindent{{\bf Figure 9:}
Frontal impact of a jet with
a dense cloud  of central coordinates (0, 0, 0) and a density 
parameter $\beta \simeq 7 \times 10^{-2}$ (the same as in the
third jet of Fig. 5). The other  
initial conditions are the same as in Fig. 1.
The times depicted are:
 $t/t_d=$ 3.5, and 6.0. 
The density lines are separated by a factor of 1.2 and
the density scale covers the range:  
$\sim$ (2$\times 10^{-1}$    -  $1.6\times 10^3$) $n_a$ (top), and
$\sim$ (2$\times 10^{-1}$    -  $1.5\times 10^3$) $n_a$ (bottom).

\noindent{{\bf Figure 10:}
Mid-plane density   and  
velocity field distribution  of an  adiabatic jet
interacting with a  cloud in an off-axis collision
after it had propagated over a distance $\sim 30 R_j$
at $t/t_d =$ 4.
The initial conditions are the same as in Figure 1.
The density scale in the contour map covers the range  
$\sim$ (4$\times 10^{-1}$ - $2.5\times 10^3$) $n_a$. 

\end{document}